\documentclass{svjour2}                    
\smartqed  
\usepackage{graphicx}
\usepackage{amssymb}
\usepackage{amsmath}
 
\newcommand{\be}{\begin{equation}}
\newcommand{\ee}{\end{equation}}
\newcommand{\bea}{\begin{eqnarray}}
\newcommand{\eea}{\end{eqnarray}}
\newcommand{\lb}{\label}
\newcommand{\bdm}{\begin{displaymath}}
\newcommand{\edm}{\end{displaymath}}
\newcommand{\D}{{\rm d}}

\begin{document}

\title{Gibbs' paradox and black-hole entropy\thanks{Dedicated to the
60th birthday of Bahram Mashhoon.}
}


\author{Claus Kiefer         \and
        Gerhard Kolland 
}


\institute{Claus Kiefer \and Gerhard Kolland 
	\at Institut f\"ur Theoretische Physik, 
Universit\"at zu K\"oln, Z\"ulpicher Stra\ss e 77, 50937 K\"oln, Germany\\
	\email{kiefer@thp.uni-koeln.de}
	\and
	Gerhard Kolland \at\email{gk@thp.uni-koeln.de}
}

\date{Received: date / Accepted: date}

\maketitle

\begin{abstract}
In statistical mechanics Gibbs' paradox is avoided if the particles of
a gas are assumed to be indistinguishable. The resulting entropy
then agrees with the empirically tested thermodynamic entropy up to a
term proportional to the logarithm of the particle number. 

We discuss here how analogous situations arise in the statistical
foundation of black-hole entropy. Depending on the underlying approach
to quantum gravity, the fundamental objects to be counted have to be
assumed indistinguishable or not in order to arrive at the
Bekenstein--Hawking entropy. We also show that the
 logarithmic corrections to this entropy, including their signs, can
 be understood along the lines of standard statistical mechanics. We
 illustrate the general concepts within the area
 quantization model of Bekenstein and Mukhanov.


\end{abstract}

\section{Introduction}
\label{intro}

Black holes are fascinating objects that have not yet revealed all
their secrets. If described by Einstein's classical theory of
relativity, they are characterized by an event horizon which encloses
a region from which nothing, not even light, can escape. If quantum
theory on a black-hole background is considered in addition, it is found that
black holes emit thermal radiation \cite{Hawking}. Black holes thus
play a key role in the search for a quantum theory of gravity
\cite{OUP}. 

Our contribution deals with the black-hole entropy and its
interpretation. We are especially interested in logarithmic
corrections to the Bekenstein--Hawking formula of black-hole entropy
and their relation to similar terms in ordinary statistical
mechanics. By highlighting their role in the discussion of Gibbs'
paradox, we give an interpretation of these terms that should
encompass all cases discussed in the literature (see \cite{Page} and
the references cited therein).

Let us, however, first give a brief introduction to this subject; more
details can be found in \cite{FN} and many other references. 
The concept of black-hole entropy first arose from formal
analogies of mechanical black-hole laws with the laws of
thermodynamics.  
The First Law of black-hole mechanics reads\footnote{Here and in 
  most of the following expressions we set $c=1$.}
\begin{equation}\label{first-law}
  \D M =\frac{\kappa}{8\pi G} \D A + \Omega_{\rm H}\D J+ \Phi
  \D q\, ,
\end{equation}
where $M$ is the black-hole mass, $A$ the area of the event horizon,
$\Omega_{\rm H}$ its angular velocity, $J$ the angular momentum,
$\Phi$ the electric potential, and $q$ the electric charge of the
black hole (if it has a charge). The quantity 
$\kappa$ denotes the surface gravity of the black hole. 
 For a Kerr black hole, $\kappa$ is explicitly given by the expression
\begin{equation}\label{kappa-explicit}
  \kappa = \frac{\sqrt{(GM)^2-a^2}}{2 G M r_+}\quad\stackrel{a \rightarrow 0}
  { \longrightarrow}\quad \frac{1}{4 G M} = \frac{GM}{R_{\rm S}^2}\, ,
\end{equation}
where 
\bdm
r_+=GM+\sqrt{(GM)^2-a^2}
\edm 
denotes the location of the event horizon.
In the Schwarz\-schild limit $a\to 0$, 
one recognizes the well-known expression for
the Newtonian gravitational acceleration. ($R_{\rm S}\equiv 2GM$ there
denotes the Schwarz\-schild radius.)

Since within the classical theory, the area $A$ of the event horizon
never decreases, this suggests a formal analogy to the Second Law of
thermodynamics, where the entropy, $S$, of a closed system never
decreases. This is re-enforced by the analogy of \eqref{first-law} with
the First Law of thermodynamics:
\be
\lb{td-first-law}
\D E=T\D S-p\D V+\mu\D N\ ;
\ee
$M$, in particular, corresponds to $E$. If we tentatively identify $S$
with a constant times $A$, the temperature should be proportional to
the surface gravity.

In the classical theory, this correspondence would remain purely formal. Its
physical significance is revealed by taking quantum theory into
account: black holes radiate with a temperature proportional to
$\hbar$, the Hawking temperature \cite{Hawking},
\begin{equation}
\label{TBH}
T_{\rm BH} =\frac{\hbar c^3}{8\pi k_{\rm B}GM} 
\approx 6.17\times 10^{-8}
 \left(\frac{M_{\odot}}{M}\right)\ {\rm K}\ .
\end{equation}
The black-hole entropy is then found from \eqref{first-law}
to read
\begin{equation}
\label{SBH}
S_{\rm BH}=\frac{k_{\rm B}c^3A}{4G\hbar}=k_{\rm B}\frac{A}{4l_{\rm P}^2}
 \ ;
\end{equation}
here, $l_{\rm P}$ denotes the Planck length,
\be
l_{\rm P} = \sqrt{\frac{\hbar G}{c^3}} \approx 1.62 \times 10^{-35} 
                                         \,{\rm m}\ .
\ee
For a Schwarzschild black hole with mass $M$ (to which we shall mostly
restrict ourselves), one has
\begin{equation}
\lb{S-Schwarzschild}
S_{\rm BH}\approx 1.07\times 10^{77}k_{\rm B}\left(\frac{M}{M_{\odot}}\right)^2
\ .
\end{equation}
In conventional units, this reads
\be
S_{\rm BH}\approx 1.5\times 10^{54}\ \frac{\rm J}{\rm K}\ 
\left(\frac{M}{M_{\odot}}\right)^2\ .
\ee
Since the entropy of the Sun is of the order of $10^{57}k_{\rm B}$, it
would experience an increase in 20 orders of magnitude in entropy
after collapsing to a black hole.\footnote{In reality, only stars with
masses bigger than about $3M_{\odot}$ collapse to a black hole.}
Gravitational collapse thus ensues an enormous increase of entropy.

It is a big challenge for any approach to quantum gravity
to provide a microscopic derivation of black-hole entropy.
The aim is to identify fundamental quantum gravitational entities
which can be counted in Boltzmann's sense to yield the entropy.
Both string theory and quantum general relativity have provided
partial answers; the fundamental entities can there be D-branes or
spin networks \cite{OUP}. The picture is, however, far from being
complete. In fact, one suffers from an embarrassment of riches, as
Steven Carlip has called it \cite{Carlip}: there are many, not
obviously related, approaches which yield the same result \eqref{SBH}.  
There thus seems to be a universal principle behind all of them, a
principle that is still veiled. 

A general mechanism which could provide such a universal principle is
connected with the notion of entanglement entropy. 
If one divides Minkowski spacetime into two different regions and
considers quantum correlations across these regions, there is a
non-vanishing entanglement entropy that is proportional to the {\em
  area} that divides these regions \cite{entanglement}.
 This result has been discussed as a support for the area law   
\eqref{SBH} in black-hole physics. But what could give the entangled
quantum degrees of freedom? Previous work uses quantum fields on a
background \cite{entanglement}. 
A universal result could perhaps be obtained from the quasi-normal
modes which are typical for the black hole itself \cite{QNM}.
These quasi-normal modes describe the characteristic perturbations of
a black hole before it reaches its final unique stationary state. No
entanglement entropy, however, has yet been calculated in this case.

An area law for the entanglement entropy is also found in analogous
situations in statistical mechanics, for example in the case of
general bosonic harmonic lattice systems \cite{CEPD}. This enforces
its universal nature.

As mentioned above, there exist various microscopic derivations of
black-hole entropy. In many of these derivations, logarithmic
corrections to \eqref{SBH} are found if one goes beyond the leading
order of the combinations. These corrections are proportional to $\ln
S_{\rm BH}$, but both the sign and the exact coefficient vary. We
shall address below these terms in more detail, but turn before to a
discussion of analogous terms in statistical mechanics.

\section{Gibbs' paradox and logarithmic corrections}
\label{sec:1}

Statistical mechanics provides a microscopic explanation of
thermodynamical relations. As has already been emphasized by 
Josiah Willard Gibbs,
one arrives at the Boltzmann entropy only when dividing the number of
permutations by $N!$, where $N$ is the number of particles. The
Boltzmann entropy coincides with the expression for the entropy found in
thermodynamics, which is known to be empirically correct. The
particles are thus counted as being indistinguishable, a procedure
that receives its justification only from quantum theory. With this
`Indistinguishability Postulate' one then gets an entropy which
is additive, at least approximately (see below). This problem in
counting states is discussed in many places, see, for example,
\cite{Zeh} and \cite{DD}.

To illustrate this situation, we
consider a system of $N$ free particles in classical statistical
mechanics. The partition sum of this model is given by
\be
\lb{Z1}
Z=\int\D^{3N}q\D^{3N}p\ \exp\left(-\frac{H}{k_{\rm
    B}T}\right)=V^N\left(CmT\right)^{3N/2}\ ,
\ee
where 
\bdm
H=\sum_{i=1}^{3N}\frac{p_i^2}{2m}
\edm
(all masses being equal), and $C$ is a constant which is independent
of the nature of the atoms.
For the entropy one gets, using standard formulae of statistical mechanics,
\bea
\lb{entropy}
S&=&k_{\rm B}\ln Z+k_{\rm B}T\frac{\partial\ln Z}{\partial T}
\nonumber\\
&=& k_{\rm B}N\left(\ln V+\frac32\ln(CmT)+\frac32\right)\ .
\eea
This expression for the entropy is {\em not} additive, that is, the entropy
does not double if volume and particle number are doubled. It would
thus be in conflict with thermodynamics.

Consider now the following consequence of this formula.
We have a box filled with an ideal gas of free particles, see
Figure~1. 

\begin{figure}[h]
  \begin{center} 
  \includegraphics[width=0.7\textwidth]{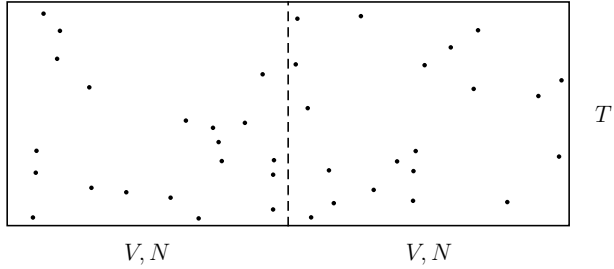} 
  \caption[]{A box with an ideal gas of particles is divided into two
    parts with equal volume and particle number.}
\end{center}
\end{figure}

A partition divides the box into two equal parts, each of
which is characterized by volume $V$ and particle number $N$. If one
removes the partition at constant temperature, one gets from
\eqref{entropy} the following increase in entropy:
\be
\lb{DeltaS1}
\Delta S =2k_{\rm B}N\ln 2\ .
\ee
On the other hand, in phenomenological thermodynamics one would expect
that $\Delta S=0$ because the situation is reversible: removing and
re-inserting the partition is a reversible process, since the state of
the gas does not change. This discrepancy is known as Gibbs' paradox
\cite{CF}. It is usually remedied by assuming that the particles are
indistinguishable and therefore 
dividing the partition sum $Z$ by
$N!$, the number of particle permutations. Instead of
\eqref{Z1} one then gets the expression
\be
\lb{Z2}
Z=\frac{V^N(CmT)^{3N/2}}{N!}\ .
\ee
In order to calculate the new entropy expression, we use Stirling's
formula, which is valid for large particle numbers $N\gg1$,
\be
\lb{stirling}
\ln N!=N\ln N-N+\frac12\ln N+\frac12\ln(2\pi)+{\mathcal
    O}\left(\frac1N\right) \ .
\ee
Instead of \eqref{entropy} we then find the expression for the `Gibbs
entropy', 
\bea
\lb{gibbsentropy}
S_{\rm Gibbs}&=&S-k_{\rm B}\ln N!\nonumber\\
&\approx& k_{\rm
  B}N\left(\ln\frac{V}{N}+\frac32\ln(CmT)+\frac52-\frac{\ln N}{2N}
  -\frac{\ln(2\pi)}{2N} \right)\ .
\eea
(This is sometimes called the `Sackur--Tetrode equation' \cite{DD}.) 
Apart from the last two terms (which are very small), this expression
for the entropy is now additive. Removing the partition in the box
described above, we now get for the change in entropy the result
\be
\lb{DeltaS2}
\Delta S\approx \frac12k_{\rm B}\ln N \ll 2k_{\rm B}N\ln 2\ ,
\ee
which, in contrast to \eqref{DeltaS1}, is almost zero. Up to a term
proportional to $\ln N$, the result of statistical mechanics now
coincides with the thermodynamical result $\Delta S=0$. 

The fact that there is not an exact coincidence can easily be
understood: the term proportional to $\ln N$ describes
fluctuations. If the partition is removed, fluctuations with larger
magnitude than in the presence of the partition
 become possible; thus, a little more states
become available. In this sense, the removal of the partition is not
quite reversible. As discussed in detail in \cite{CF}, this situation
corresponds to a `microscopic preparation', where $N$ identical
particles are initially placed on each side of the partition at the
same temperature. If, instead, one makes a `macroscopic preparation'
(with knowledge only about the pressure and the temperature, but with 
no information about the exact value of $N$), one finds the
exact result $\Delta S=0$ upon removing the partition. 

It is important to emphasize in this connection the important
difference between identity and indistinguishability \cite{Zeh}. 
In classical mechanics, different particles are not identical even if
they are indistinguishable; in principle, they can be identified and
have therefore to be counted separately.\footnote{To quote from Otto
  Stern's paper \cite{Stern}: ``The conception of atoms as particles
  losing their identity cannot be introduced into the classical theory
without contradiction. This is possible only on the ground of the
non-classical ideas of quantum theory.''}
In quantum theory, on the other hand, one
does not have `particles', but only field modes. If one has, for
example, a wave packet with two bumps, the exchange of the bumps
describes the same state -- in this sense both states (before and
after the exchange) are identical. It is only this identity that
justifies the division of the partition sum by $N!$.

Quite generally, one can write the ensemble entropy related to a
reduced density operator as a sum of the averaged physical entropy
(which is a definite function of volume, temperature, etc.)
plus the entropy of missing information, the latter being usually much
smaller than the former \cite{Zeh}. Consider, for example, the case of
a grand-canonical ensemble, for which the density operator reads
\be
\rho=\frac1Z\exp\left(-\frac{H-\mu N}{k_{\rm B}T}\right)\ ,
\ee
where $\mu$ is the chemical potential.
Only the mean particle number is specified here, because the system is
assumed to be in contact with a particle reservoir. If this contact is
closed, the particle number assumes a definite value, but this value
is unknown. The corresponding relative entropy of 
missing information about this value is of order $\ln
N/N$, which just corresponds to the entropy increase in \eqref{DeltaS2}. 

\newpage

To conclude this section, we briefly discuss a simple model which can
also serve as an analogy for the black-hole case. Consider a set of
$N$ spin-1/2 particles out of which $n$ point up and $N-n$ point
down:

\begin{figure}[h]
  \begin{center} 
  \includegraphics[width=0.5\textwidth]{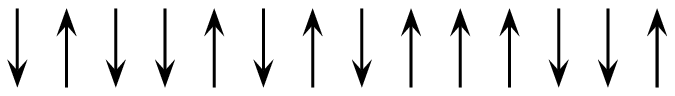} 
 \end{center}
\end{figure}

Since $N$ is assumed to be given, we have a microcanonical
ensemble. We define the entropy as the logarithm of the number of
configurations with $n$ spins up and $N-n$ spins down,
\be
\lb{spinentropy}
S=\ln \left(\begin{array}{c} N\\ N-n \end{array}\right)=
\ln\left(\begin{array}{c} N\\ n\end{array}\right)\ .
\ee
In a realistic setting, $n$ could correspond to the magnetization as
the given macroscopic quantity. 

Consider first the `equilibrium case' $n=N/2$.
Using \eqref{stirling}, one gets from \eqref{spinentropy},
neglecting terms of order $1/N$,
\be
\lb{equilibrium}
S=N\ln 2-\frac12\ln N+\frac12\ln\frac{2}{\pi}\ .
\ee
Defining $S_0\equiv N\ln 2$, we see that the relative contribution of
the second term is just of order $\ln N/N$. In contrast to the above,
it comes with a minus sign; the reason is that it does not describe
missing information because $N$ is fixed 
from the very beginning (since we have here a
microcanonical ensemble). Note that we can approximately write 
\bdm
S\approx S_0-\frac12 \ln S_0\ .
\edm
In the general case \eqref{spinentropy} we get (assuming both $n$ and
$N$ to be large numbers)
\be
\lb{nonequilibrium}
S=-N(w\ln w+(1-w)\ln(1-w))-\frac12\ln(Nw(1-w))-\frac12\ln(2\pi)\ ,
\ee
where $w=n/N$. Defining now
\be
\lb{entropy2}
S_0\equiv -N\left(w\ln w+(1-w)\ln(1-w)\right)\ ,
\ee
we find
\be
\lb{entropy3}
S=S_0-\frac12\ln S_0-\frac12\ln\left(\frac{2\pi w(1-w)}{\alpha}\right)\ ,
\ee
where $\alpha=(w-1)\ln(1-w)-w\ln w$. 

\begin{figure}[h]
  \begin{center} 
  \includegraphics[width=1.0\textwidth]{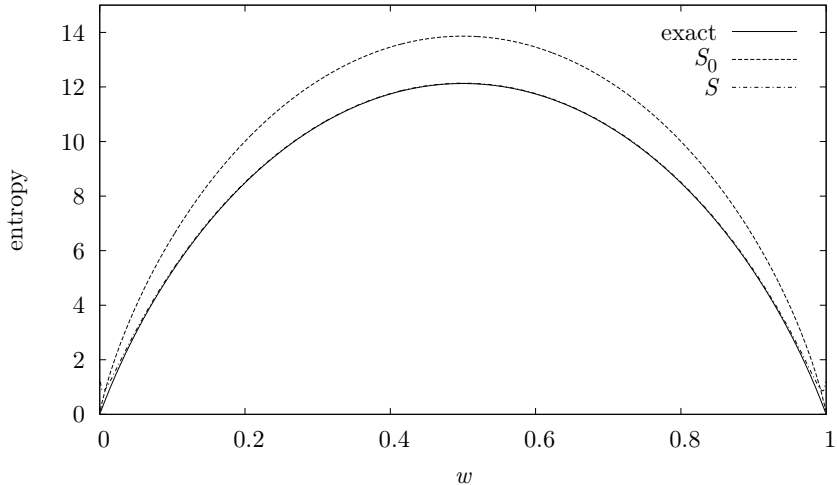} 
  \caption[]{Comparison of the exact entropy \eqref{spinentropy} 
    with the approximations
    with, Eq. \eqref{entropy3},
and without, Eq. \eqref{entropy2}, the logarithmic correction term for
$N=20$.
 The difference between the exact expression and \eqref{entropy3} is
 hardly noticeable.}
\end{center}
\end{figure}

Figure~2 compares the exact expression for the entropy with $S_0$ and
$S$ according to \eqref{nonequilibrium}. One easily sees that 
\eqref{nonequilibrium} is an excellent approximation unless $n$ or $N$
are small numbers. 

\section{Logarithmic corrections to black-hole entropy}
\label{sec:2}

The big challenge in understanding black-hole entropy is to provide a
microscopic interpretation for it. This is possible only in quantum
gravity, a theory which presently does not exist in a complete
form. Two major approaches within which the interpretation of the
entropy can be tackled are quantum general relativity and string
theory \cite{OUP}. 
For example, loop quantum gravity, which is a particular case of
quantum general relativity, 
gives a discrete spectrum for an appropriately defined area operator;
the area of the event horizon can then only assume discrete values
\cite{OUP,Rovelli}. Before we turn to this case, we demonstrate the essential
features in the context of a much simpler model:    
we assume the presence of an equidistant area spectrum as put forward
by Bekenstein and Mukhanov \cite{BM}. Such a spectrum can also be
found from quantum geomtrodynamics \cite{VW}. It is given by
\be
\lb{areaspectrum}
A_N=(4\ln k)l_{\rm P}^2N\ ,
\ee
where $k$ is an integer number $> 1$. The intuitive picture is
that the horizon is divided into small cells with area 
$(4\ln k)l_{\rm P}^2$, see Figure 3.

\begin{figure}[h]
  \begin{center} 
  \includegraphics[width=0.6\textwidth]{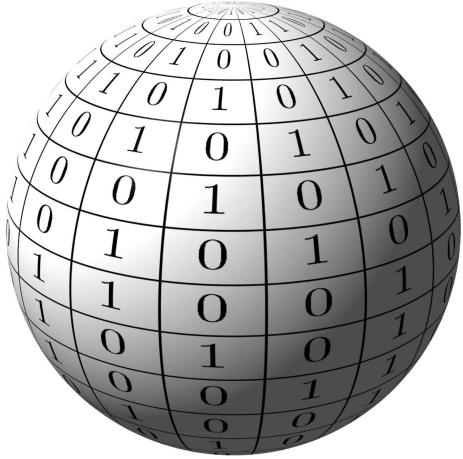} 
  \caption[]{Symbolic attachment of bits to the surface of a black hole.}
\end{center}
\end{figure}

In each cell there are `spins' which can assume $k$ different
values. The simplest case is $k=2$, so one bit of information can be
put on each cell, cf. Wheeler's notion of `it from bit' \cite{Wheeler}.
 Inspecting the spin model discussed at the end of
the last section, one recognizes that one can identify the
Bekenstein--Hawking entropy \eqref{SBH} with 
the leading term $S_0=N\ln 2$ of the equilibrium entropy. 
This is not possible for the non-equilibrium case
\eqref{nonequilibrium}, which sounds reasonable, since one would
expect that the spins are equally distributed on the surface of a big
black hole. Taking into account the corrections to $S_0$ from
\eqref{equilibrium}, one arrives at the following expression for the
black-hole entropy in the Bekenstein--Mukhanov model when using
\eqref{areaspectrum}:
\be
\lb{approximateentropy}
S=\frac{A_N}{4l_{\rm P}^2}-\frac12\ln\frac{A_N}{4l_{\rm P}^2}+
\frac12\ln\frac{2}{\pi}+\frac12\ln(\ln 2)\ . 
\ee
The logarithmic correction term has a negative sign because one has
here a microcanonical ensemble -- the value of the area is {\em
  fixed}, and one therefore has a slight increase of information from
the knowledge of the microstate compared to \eqref{SBH}. The situation
would be different in a grand-canonical setting in which
the area can fluctuate and only the mean
value of $A$ is known; then the sign of the logarithmic term would be
positive, corresponding to missing information. This difference has
been clearly emphasized by Gour \cite{Gour}.

The analogue of the black-hole mass $M$ is in statistical
mechanics the energy $E$; the analogue of the area $A$ is the particle
number $N$. Situations in which $A$ is fixed lead to an increase of
information by going beyond the highest order in $N$, whereas
situations in which $A$ fluctuates lead to a decrease of information
(of information about the exact value of $A$).

In this simple model we can also evaluate the exact value for the
entropy by making use of \eqref{spinentropy}. Inserting there
$N=A_N/4l_{\rm P}^2\ln 2$ and $n=N/2$, one gets
\be
\lb{Sexact}
S=\ln\frac{\left(\frac{A_N}{4l_{\rm P}^2\ln 2}\right)!}
{\left[\left(\frac{A_N}{8l_{\rm P}^2\ln 2}\right)!\right]^2}\ .
\ee
It may be of interest to compare this exact expression with the
approximate expression \eqref{approximateentropy} in order to see how
good the approximation is.
Consider, for example, a small black hole with $N=20$. We then have
$A_{20}\approx 55.45\ l_{\rm P}^2$, and the number of states is 
\bdm
\left(\begin{array}{c} 20\\10\end{array}\right)=184756\ ,
\edm
which corresponds to the exact entropy $S=\ln 184756\approx 12.127$. 
 The approximate entropy, as found from \eqref{approximateentropy}, is
 $S \approx 12.139$, which is only slightly larger than the exact
 value. (The dominant term $A/4l_{\rm P}^2$ is about $13.86$.)
Thus, although this black hole is rather small, the
 approximation found by using Stirling's formula is still quite
 good. This black hole has a radius $R_{\rm S}\approx 2.1\ l_{\rm P} $,
 a mass $M\approx 1.05\ m_{\rm P}$, 
where $m_{\rm P}=\hbar/l_{\rm P}$ is the Planck mass 
and, from \eqref{TBH}, a temperature 
$T_{\rm BH}\approx 4.7\times 10^{17}$ GeV. Once the black hole
approaches the Planck scale, the whole approximation breaks down and
one would have to use the full quantum theory of gravity \cite{OUP}. 

As we have seen, the relative contribution of the logarithmic
correction term is negligible even for relatively small black holes. 
Even a primordial black
hole with $M\approx 10^{-18}M_{\odot}$ 
(which could have been formed in the early universe) gives a logarithmic
correction with relative contribution only of $4.4\times 10^{-40}$. 

In the Bekenstein--Mukhanov model, logarithmic contributions to the
main contribution to the entropy appear naturally, see
\eqref{approximateentropy}. Such terms also arise from various
approaches to quantum gravity \cite{Page}. Let us concentrate on two
of them: loop quantum gravity and string theory.

In loop quantum gravity, black-hole entropy follows from counting all
possible punctures of a spin network with the horizon. A spin network
is characterized by a collection of quantum numbers $j$ and $m$, where
$j\in\{\frac12,1,\frac32,\ldots\}$ and $m=-j,\ldots,j$. 
The combinatorial problem is difficult \cite{DL,GM}. 
The entropy turns out to be proportional to the area only if the
exchange of nodes in the spin network produces a {\em different}
state, that is, if the counting is performed without dividing by
the corresponding number of permutations. Otherwise the entropy would
come out to be proportional to the {\em square root} of the area
instead of the area itself. This would be in conflict with the laws of
black-hole mechanics, which correspond to the level of
thermodynamics. If the nodes are treated as distinguishable, the
proportionality to the area is found. The exact expression \eqref{SBH} 
can only be recovered if an unknown
parameter of loop quantum gravity (the Barbero--Immirzi parameter
$\beta$) is chosen appropriately. A logarithmic correction term arises
if one imposes in addition a `spin projection constraint' of the form
$\sum_im_i=0$. It
turns out to be of the same form as in \eqref{approximateentropy},
that is, it comes with a factor $-1/2$. As we have discussed above,
the reason for the minus sign is the fact that the area of the horizon
is assumed to be fixed in this approach and that the additional
constraint therefore can only lead to an increase of information
(decrease of entropy), different from the case where the area
fluctuates. 

In string theory, the situation is different. There, the Bekenstein--Hawking
entropy \eqref{SBH}
is recovered by counting states of D-branes in a weak-field
situation without black holes but duality-related to a situation with
black holes, see, for example, \cite{Mohaupt} and the references
therein. Corrections are also found, and they start in many cases with
a term proportional to $\ln A$. The signs of these terms vary, which
again seems to depend on whether area is fixed or not. In contrast to
loop quantum gravity, invariance under permutations is assumed, that
is, the fundamental `particles' are assumed to be indistinguishable. 

As the analogy with the above examples shows, the terms proportional
to $\ln N$ come mainly into play through the application of Stirling's
formula beyond the highest order. They are thus not necessarily of
`quantum origin', but can arise already from classical statistical
mechanics (as can be seen from the fact that they are not proportional
to $\hbar$). 

In most of the above foundations of black-hole entropy, the black hole
is considered to be in a state corresponding to a microcanonical
ensemble. The canonical ensemble (black hole in a heat bath) is
undefined in an asymptotically flat spacetime because it would yield a
negative specific heat and formal energy fluctuations with a negative
variance. Therefore, logarithmic corrections cannot be computed in
this case. The situation improves 
if the black hole is put in a box \cite{AD} or
in anti-de~Sitter spacetime \cite{Page}. As discussed in detail by Don
Page, logarithmic terms can easily show up by going from a
microcanonical to a canonical ensemble and vice versa. This shows that
``entropies need to be defined carefully before there is any
unambiguous meaning to logarithmic corrections'' \cite{Page}. 

We conclude this section by presenting some numerical examples for the size of
the logarithmic corrections. We assume that 
we have the relation (in units of $k_{\rm B}$)
\be
\lb{gbh}
S=S_{\rm BH}-\frac12\ln S_{\rm BH}+ \ldots \ . 
\ee
Let us consider, for example, the galactic black hole, 
which lurks in the centre of our Milky Way and which has a
mass $M\approx 3.6\times 10^{6}M_{\odot}$ \cite{Eckart}. From
\eqref{S-Schwarzschild} one gets
\be
S_{\rm BH}\approx 2\times 10^{67}\ \frac{\rm J}{\rm K}\ ,
\ee
which is, of course, enormous compared to any laboratory-scale
entropy. It is even bigger than the entropy of the cosmic background
radiation, which is known to dominate the non-gravitational entropy of
the observable part of our Universe \cite{Zeh}. The galactic black
hole also possesses angular momentum, which slightly reduces its
entropy (in the extremal case, one would have half of the value in
\eqref{gbh}). 
The logarithmic correction leads to the
following tiny decrease in entropy:
\be
-\frac12\ln S_{\rm BH}\approx -1.4\times 10^{-21}\ \frac{\rm J}{\rm K}\ , 
\ee
which is about $7\times 10^{-89}$ of $S_{\rm BH}$. This would be a
negligible number even for laboratory scales!

\section{Interpretation and conclusion}
\label{sec:3}

We have seen that various conceptual issues that arise in the counting of
microscopic states for the black hole are fully analogous to ordinary
statistical mechanics. We have shown, in particular, that logarithmic
corrections to the Bekenstein--Hawking area law occur in a natural
way. These corrections are not a priori of quantum nature, but have
their origin in combinatorial relations such as Stirling's
formula. The sign of a logarithmic term can be understood as either
related to missing information (if it is positive) or increase of
information (if it is negative), depending on whether the horizon area
is fixed or not. 

The traditional Gibbs paradox has been resolved by assuming that the
microscopic particles are identical. While in the classical theory
this is an ad hoc assumption without justification, it can be
understood from quantum theory, which does not contain fundamental
particles. This shows that one should get rid of classical pictures as
much as possible \cite{Zeh}. 

In the case of black holes, the Bekenstein--Hawking area law
\eqref{SBH} plays in a certain sense the role of the entropy
expression in thermodynamics; microscopic derivations from quantum
gravity are expected to recover it at leading order. It is therefore
of interest to see whether a new type of Gibbs paradox may
arise. Surprisingly, the situation seems to be opposite in loop
quantum gravity and in string theory: whereas the former needs
fundamental entities that are distinguishable, the latter works with
indistinguishable structures in order to recover \eqref{SBH}. In
analogy with quantum theory one would have expected that the
fundamental `particles' are identical, so the situation in loop
quantum gravity needs perhaps some further understanding to become
intuitive.   

Black holes are open systems. They can thus only be understood if their
interaction with other degrees of freedom are consistently taken into
account \cite{info}.
 In quantum theory, this is known to lead to the emergence of
classical properties through decoherence \cite{deco}.
In a similar way, the black hole, which is fundamentally described by
quantum theory, should assume classical properties by interacting with
other fields: these could be additional quantum fields or the quantum
perturbations (the quasi-normal modes) of the black hole itself.
The thermal nature of Hawking radiation can, for example, be
understood as arising from decoherence \cite{CK01}. 
The quantum entanglement between these other fields and the quantum
gravitational states of the black hole could be at the heart of the
black-hole entropy. The corresponding calculation should automatically
avoid Gibbs' paradox and lead to further insight into the
interpretation of the underlying quantum theory of gravity.

\begin{acknowledgements}
C. K. thanks Thomas Mohaupt and H.-Dieter Zeh for useful discussions.
\end{acknowledgements}


\end{document}